# Defects induced ferromagnetism in Mn doped ZnO

S. Chattopadhyay[a], S. K. Neogi[a], A. Sarkar[b], M. D. Mukadam[c], S. M. Yusuf[c],A. Banerjee[a], S. Bandyopadhyay[a*]

[a]*Department of Physics, University of Calcutta, 92 A P C Road, Kolkata-700009, India*
[b]*Department of Physics, Bangabasi Morning College, Kolkata -700009, India*
[c]*Solid State Physics Division, Bhaba Atomic Research Centre, Mumbai-400085, India*

## Abstract

Single phase Mn doped (2 at %) ZnO samples have been synthesized by solid-state reaction technique. Before the final sintering at 500 $^{o}$C, the mixed powders have been milled for different milling periods (6, 24, 48 and 96 hours). The grain sizes of the samples are very close to each other (~ 32 ± 4 nm). However, the defective state of the samples is different from each other as manifested from the variation of magnetic properties and electrical resistivity with milling time. All the samples have been found to be ferromagnetic with clear hysteresis loops at room temperature. The maximum value for saturation magnetization (0.11 $\mu_B$/ Mn atom) was achieved for 96 hours milled sample. Electrical resistivity has been found to increase with increasing milling time. The most resistive sample bears the largest saturation magnetization. Variation of average positron lifetime with milling time bears a close similarity with that of the saturation magnetization. This indicates the key role played by open volume vacancy defects, presumably zinc vacancies near grain surfaces, in inducing ferromagnetic order in Mn doped ZnO. To attain optimum defect configuration favorable for ferromagnetism in this kind of samples proper choice of milling period and annealing conditions is required.

* Author to whom correspondence should be addressed; Fax: +91-33-23519755, e-mail: sbaphy@caluniv.ac.in

## 1. Introduction

In the last couple of years, dilute magnetic semiconductor (DMS) systems have drawn enormous interest in the scientific community [1-12]. This is due to their potential for application in spintronic devices. The euphoria acquired momentum following the prediction by Dietl et al. [13] that diamagnetic ZnO can be made ferromagnetic by doping with transition metal (TM) ions such as Mn. The exact nature of the ferromagnetic coupling of spins in such materials is a matter of debate till date. A number of new theoretical approaches [14] have been developed to describe DMS systems; very often such theories contradict each other in their basic assumptions and final conclusions. On experimental side too, claims and counterclaims of ferromagnetism (FM) in doped or un-doped ZnO are being published frequently [3,15]. However, one important understanding has evolved that lattice defects play crucial role in stabilizing the FM state in DMS systems. Some type of defects or disorder favors ferromagnetism and others competes with this phenomenon. Reports in favor of Zn vacancy ($V_{Zn}$) [8,16], O vacancy ($V_O$) [17,18] or Zn interstitial ($I_{Zn}$) [19] mediated ferromagnetism can be found. It is also found that the defect species involved in inducing ferromagnetism in one ZnO based DMS system may not be effective in similar others [4,7,20]. The site occupied by the dopant atom [16], its size, electronic configuration and defect decoration [9], lattice strain [21], band gap [5] etc. determines the final and modified band structure of the ZnO host lattice. A delicate balance of several defect types is, thus, crucial for giving rise to net ferromagnetic response of the system. Not only point defects, but the role of extended defects such as grain boundaries (GB) or dislocations in a polycrystalline material is important in this context. These defects have been proved to be helpful [7,22] to reach the required defect structure for ferromagnetic interaction. In this study, we aim to probe the role of defects on magnetic properties of polycrystalline ZnO doped with low concentration of Mn (2 at%). The main focus is the search for a suitable defect related parameter on which the net magnetization of the material depends.

A variety of defects is most likely in polycrystalline or nanocrystalline materials and they reside near the grain surface regions. Several recent reports identify the defects in the vicinity of the grain surfaces are responsible for the intrinsic FM in TM doped ZnO [7,9,15,22,23]. More specifically, it has been proposed that only a thin nanometric region surrounding the grains become ferromagnetic with suitable defect structure [4,7]. Positron annihilation lifetime (PAL) technique is one of the best techniques for probing vacancy defects in semiconductors [20,24,25]. For systems with nanometer size length scale, where positron diffusion length in the material is higher than the mean grain size, positrons populate near the disordered grain boundaries and annihilate there [26]. The resultant PAL parameters bear information regarding the nature and abundance of defects at the annihilation sites. However, PAL is very much sensitive for open volume defects only. In ZnO, $V_{Zn}$ related defects/defect complexes or their clusters are the major trapping sites of positrons [24,25,27]. The origin of ferromagnetism in TM doped ZnO, if it is related to the open volume defects near the GB, the magnetic response should scale with the PAL parameters. Recently, Khalid et al. [8] have employed positron annihilation spectroscopy on nitrogen doped ZnO samples and found [8] that the ferromagnetism is related with the size of the open volume defect clusters. In un-doped ZnO nanocrystals, Wang et al. have found [27] that ferromagnetism disappears when the signature of the $V_{Zn}$ related defects in the PAL spectrum is reduced. Another recent work has also pointed out [12] the role of open volume defects in inducing ferromagnetism in Li doped ZnO.

## 2. Experiment And Data Analysis

Mn doped ZnO pellet samples have been synthesized by solid-state reaction technique [2,28]. High purity ZnO (99.99%, Sigma-Aldrich, Germany) and $MnO_2$ (99.9%, Sigma-Aldrich, Germany) have been taken in stoichiometric amounts in an agate container. Four steps have been carried out for sample synthesis. In the first step, the sample has been milled for one third of the total milling time in a "Fritsch planetary mono mill" using agate balls. In the next step, the milled powder has been sintered at 400 $^{o}$C for 8 hours in air. The sintered material has been milled again for the remaining two third of the total milling time. Finally, the powder have been made pallets and re-sintered at 500 $^{o}$C for 12 hour in air. The final annealing temperature was chosen to be 500$^{o}$C in the light of earlier reports that the samples annealed at 500 $^{o}$C shows RT FM [1,2,28] and above 500$^{o}$C magnetization decreases [1,2,10]. The total milling time has been varied from sample to sample as 6, 24, 48 and 96 hours. The ball to mass ratio was maintained at 1:1 throughout the milling process. A drastic lowering of grain size is not expected with such a ball to mass ratio. Increase of overall disorder and improved site occupancy of the dopants (here Mn) are expected with higher milling time. It has also been observed earlier that appearance of impurity phase started from 4 at% of Mn doping in ZnO samples [28]. So in the present study, the doping level of Mn has been kept 2 at % to avoid the formation of any types of impurity phase. The phase characterization was performed by X-ray diffraction (XRD) measurements [Philips, Model: PW1830] with Cu $K_{\alpha}$ radiation. RT resistivity of these highly resistive samples was carried out using conventional two-probe technique using Kithlay 6514 electrometer. The magnetization measurements of the samples at RT were performed by a 12 Tesla vibrating sample magnetometer (VSM) [Oxford Instruments]. For PAL study, a 10-μCi $^{22}$Na positron source was sandwiched between two identical plane faced pallets of the samples. The PAL spectra were measured with a fast-slow coincidence assembly having 182±1 ps time resolution [26]. Measured spectra were freely fitted by computer program (PATFIT-88) to obtain the possible lifetime components $\tau_i$, and their corresponding intensities $I_i$. In nanomaterials, wide distribution in the size of the open volume defects exists. So assignment of the defect types with lifetime components is not trivial due to limited accuracy of the least square fitting procedure of the PAL spectrum of finite statistics. In this situation, it is better to choose statistically more accurate parameter [25], the average lifetime, $\tau_{av}$ (= $\sum\tau_i I_i/\sum I_i$). The change in $\tau_{av}$ is related to the modification of overall defect state of the system.

## 3. Results and Discussion

XRD patterns of 2 at% Mn doped ZnO samples synthesized at 6, 24, 48 and 96 hours of milling are shown in Fig. 1. The indices in the spectra, as shown in Fig. 1 clearly indicate the expected positions of the peaks for the wurtzite crystal structure of ZnO and no signature of impurity peaks have been observed. Any impurity phase generated below the detection limit of XRD can't originate RT FM, as none of these possible phases (e.g. $ZnMn_2O_4$, $Mn_3O_4$ etc) is FM at RT [16,29]. As the annealing temperature is same for all the samples (milled for different time), the resultant grain size is expected to be more or less same. This can be seen from the enlarged view of the (002) peak FWHM (peak width at half maximum) as observed from the inset of Fig 1. However, three competing processes can alter the defective nature of the grain boundaries/grain interiors even if the grain sizes of the samples are more or less same. It is to be noted that the final defective state (open volume defects and the site occupancy of Mn atoms altogether) of the GB depends on the local free energy minimization. Mechanical milling induces structural disorder in the GB as well in the grain interiors. A fraction of defects at the grain interiors may migrate [26] to GB region due to annealing. A small fraction also gets recovered [26,27] with annealing at 500 $^{o}$C. As a whole, the (002) XRD peak shows a monotonic shift in the lower 2θ direction with higher milling time (inset of Fig 1). It is probably a reflection of more grain surface disorder and improved occupancy of Mn atoms at the Zn sites [30]. Both of these factors can contribute to the enhancement [30] of resistivity with increased milling time (Table 1). Very high resistivity of the samples indicates that carriers are localized. So it is consistent [3] that the sample with highest resistivity shows highest ferromagnetic ordering in application of magnetic field.

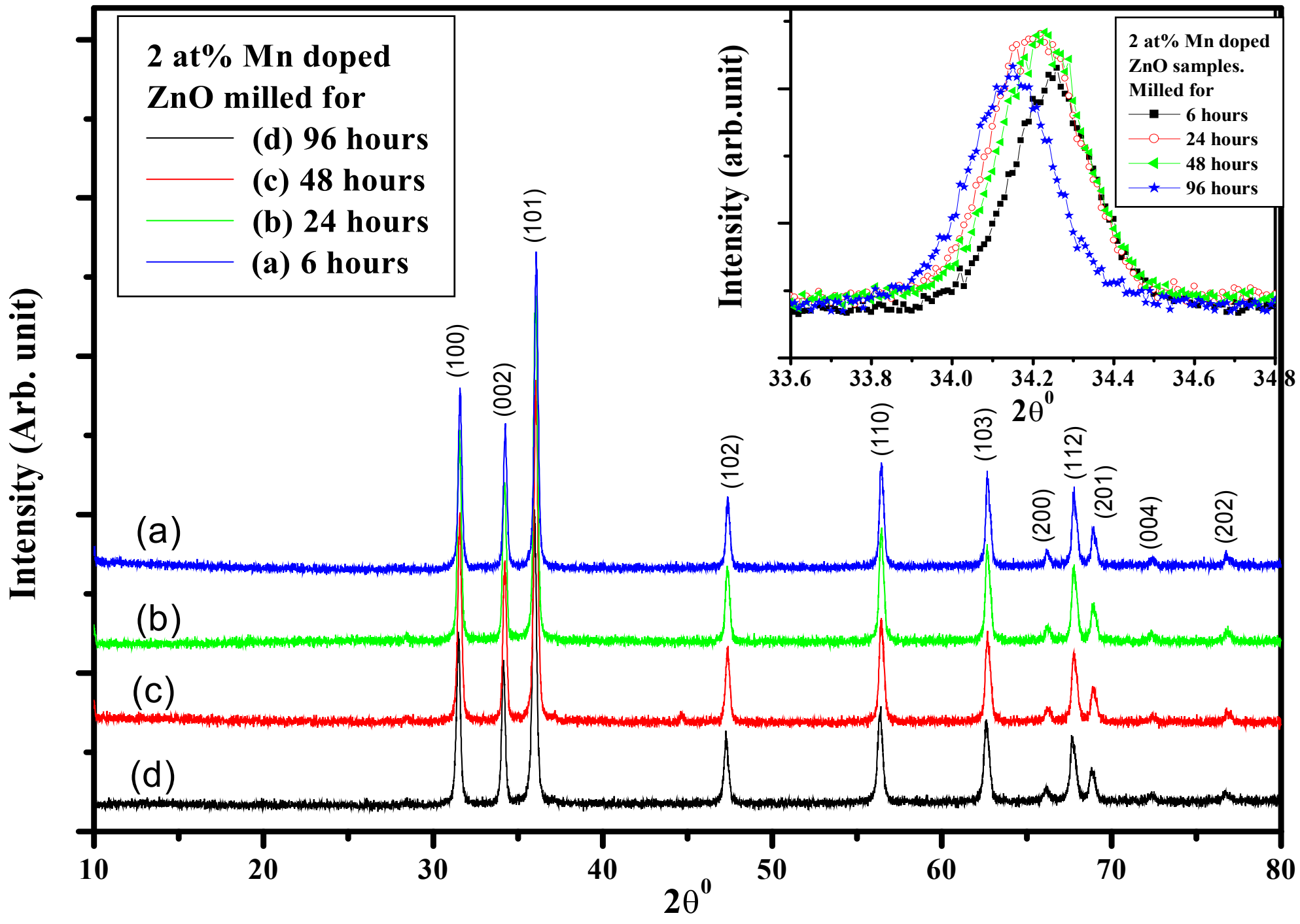

Fig. 1: XRD Spectra of (a) 6, (b) 24, (c) 48, and (d) 96 hours milled Zn(Mn)O samples. Inset: Enlarged view of the 002 diffraction peak region.

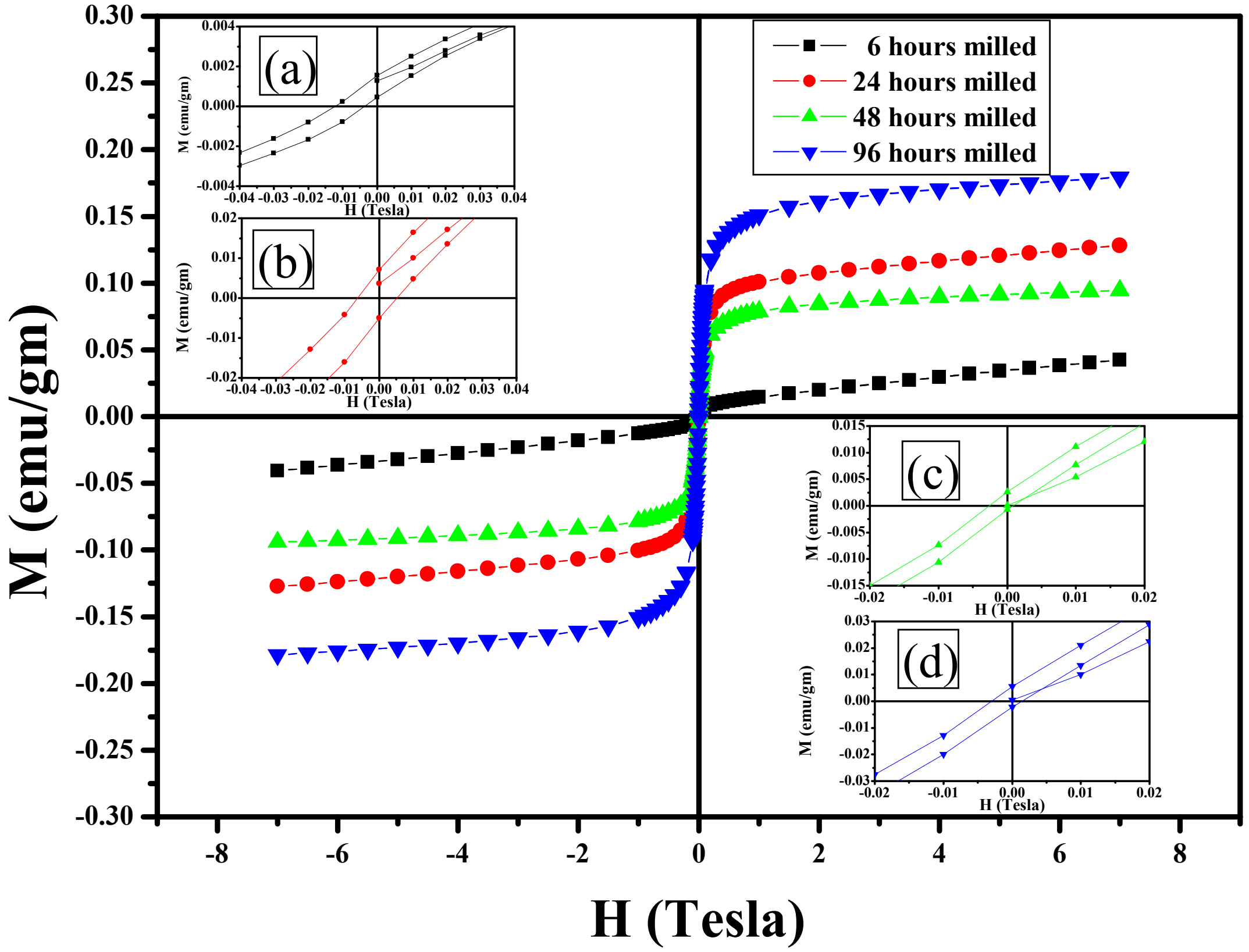

Fig. 2: Field dependence of magnetization at room temperature for 2 at% Mn doped ZnO samples milled for (a) 6, (b) 24, (c) 48 and (d) 96 hours. Inset of (a), (b), (c), and (d) indicates the enlarged portion of the *M-H* curve to clearly visualize the ferromagnetic hysteresis loop for 6, 24, 48 and 96 hours milled 2 at% Mn doped ZnO samples respectively.

The field dependence of dc magnetization (M) at RT for all the samples is plotted in Fig. 2. All the M-H curves exhibit FM behavior. The insets in Fig. 2 clearly demonstrate the FM nature of the samples at room temperature. The variations of saturation magnetization ($M_S$), remanence ($M_R$) and coercivity ($H_C$) with milling time have been presented in Fig. 3. The variation of such parameters with milling time indicates an important fact. Although the grain sizes are nearly same (~ 32 ± 4 nm) for all the samples but the magnetic domain lengths or the domain wall widths have been effectively varied from sample to sample. The nature of variations of $M_S$ and $M_R$ with milling time shows similar trend. An overall increasing tendency (for $M_S$ and $M_R$) for 6 to 96 hours of milling with slight decrease for 48 hours milled sample has been observed. Also to note, for 6 hour milled sample the magnetization has not been completely saturated; though for samples with higher milling times, the saturation tendencies of magnetization are better. We feel that a competing paramagnetic tendency is dominant for 6 hours milled sample. So the actual value of $M_S$ for 6 hours milled sample is much lower and paramagnetic contribution is superimposed on it. The maximum value for $M_S$ (0.11 $\mu_B$/Mn atom) was achieved for the 96 hours milled sample. The value is almost comparable with some of the reported results [7,31]. But it is much smaller compared to the theoretical value of $Mn^{+2}$ state (5 $\mu_B$/Mn atom) [13], antiferromagnetic (AFM) coupling between neighboring Mn atoms [17,32] might reduce the strength of FM. Following the argument of Mukherjee et al. [32], we can estimate that only ~ 2% of the total $Mn^{+2}$ ions are aligned ferromagnetically at room temperature for 96 hours milled sample (for other samples the percentage is even less). Remaining Mn ions are randomly distributed mostly at interstitial sites. Except Mn-Mn nearest neighbor AFM interaction (inside insulating matrix), there may exist isolated Mn atoms which are paramagnetic in nature. Such paramagnetic contribution is largest for 6 hours milled sample. Increasing the milling time further, more substitutional Mn sites ($Mn^{+2}$) or $Mn^{+2}$– defect pairs are created [11,18] giving rise to enhanced FM character of the samples. On the other hand, the coercivity values decrease when the milling time increases from 6 to 48 hours and finally it saturates. Our result is consistent with other reports where $M_S$ and $H_C$ show opposite trends [31,33] which are typical for a soft ferromagnet. There exist other reports [18,32] where $M_S$ and $H_C$ change in a similar fashion. It is expected that both the features have defect related origin. It is possible that two different mechanism [3] of magnetic order (depending on the nature of defect) in TM doped ZnO are responsible for distinct features of $M_S$ and $H_C$ variation.

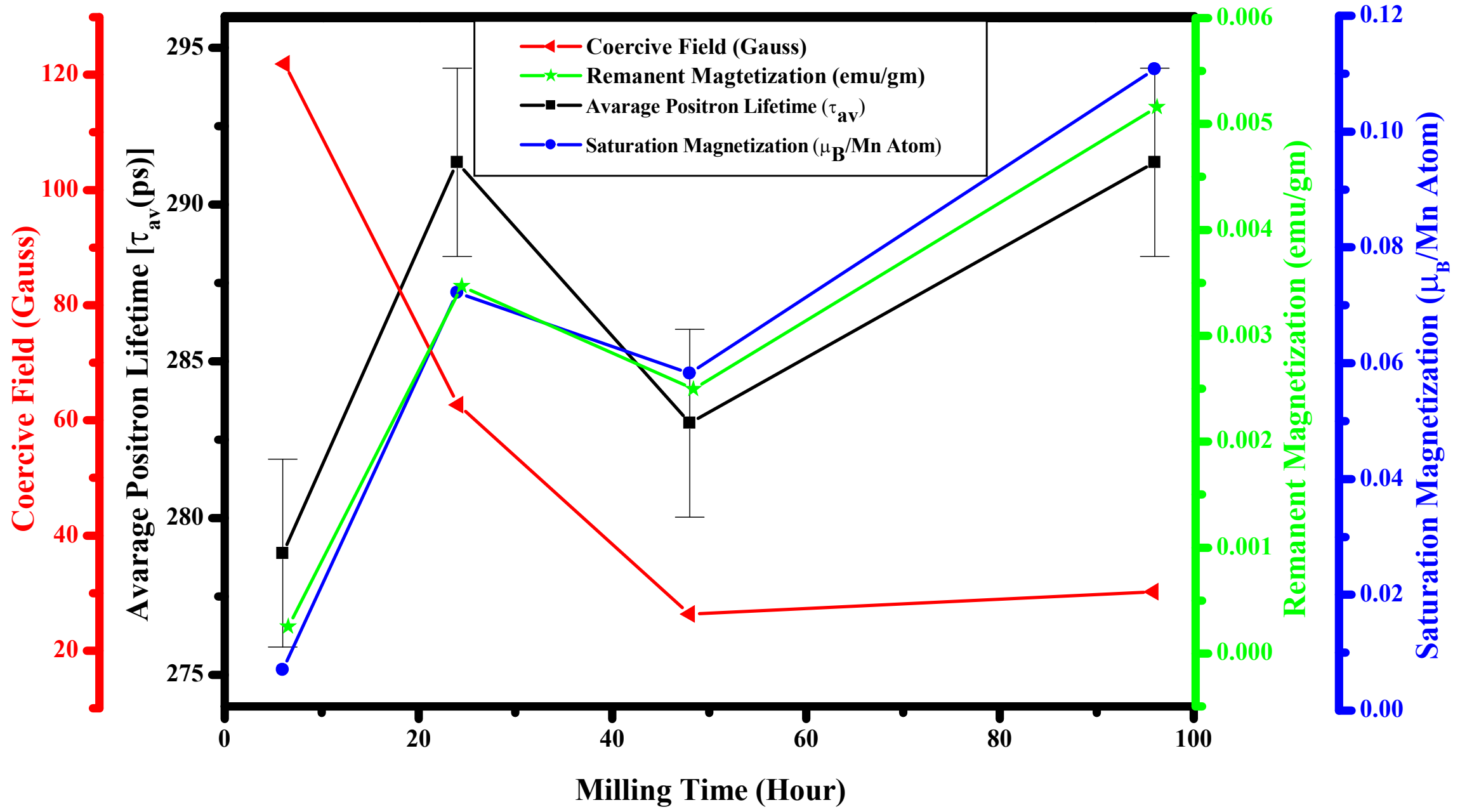

Fig. 3: Variation of saturation magnetization, remanence, coercive field and fitting independent positron annihilation lifetime parameter $\tau_{av}$ with milling time of the samples.

The results of the PAL spectrum analysis are shown in table 1. The most important lifetime component is $\tau_2$, which arises from vacancies/vacancy clusters in the material and represents the spatial extension (size) of the open volume to some extent. The corresponding intensity ($I_2$) indicates the abundance of the related defect species. In ZnO, $V_{Zn}$ is the major trapping site and the value of $\tau_2$ for isolated $V_{Zn}$ in ZnO is ~ 230 ps [24]. In comparison, such a high value of $\tau_2$ (here ~ 330 ps), generally found in granular ZnO [25], can be attributed to 7-8 agglomerated vacancies of $V_{Zn}$-$V_O$ divacancy type [34]. Such a highly defective region can produce a foam like [7] network at the GB, the nature of which is modified with milling and subsequent annealing. In the doped system, a partial modification of such vacancy clusters cannot be ruled out [35]. Still it can be understood that the dominant defect species in the present samples must be $V_{Zn}$ related clusters or complexes as evident from the sensitivity of $\tau_2$ and $I_2$ with milling time. Our measured values of $\tau_2$ and $I_2$ are also consistent with a recent report [35] on PAL spectroscopic investigation of $Zn_{1-x}Mn_xO$ samples. However, we will focus our discussion on $\tau_{av}$, which is free from fitting related errors, if any. The variation of $\tau_{av}$ with milling time is also shown in Fig. 3. The most striking feature is that the variation follows a similar trend as $M_S$ and $M_R$. This indicates that the disorder which PAL spectroscopy probes, controls $M_S$ and $M_R$. It has been reported that RT FM in Mn doped ZnO samples arise particularly from the $V_{Zn}$ along with substitutional Mn ($Mn_{Zn}$) sites i.e. $Mn_{Zn}$+$V_{Zn}$ pair formation in the system [16] as it energetically favors FM as compared with AFM interaction. Magnetization should increase with increasing number of such defect species which in turn increases $\tau_{av}$. So we strongly recommend that $V_{Zn}$ play the determining role in mediating FM for Mn doped ZnO. As we mentioned earlier, the value of $\tau_2$ is quite high in comparison to the isolated zinc vacancies in single crystalline ZnO. The parameter, $\tau_2/\tau_B$ ($\tau_B$ is the bulk lifetime of positrons in ZnO) is thought to be an indicator of the size of the vacancy clusters present in the sample. For the sample with highest $M_S$, the value of $\tau_2/\tau_B$ becomes 2.2 (taking $\tau_B$ = 181 ps) [24,27], a further indication of positron annihilation at large vacancy clusters. So the presence of extended defects [22] or foam like network [7] at the grain boundaries may serve as the background disorder for giving rise to FM interaction between particular defect species. A recent report has also revealed [36] that increased grain surface disorder in ZnO nanorods, estimated from the ratio of defect level and exciton luminescence, can be responsible for enhancement of $M_S$. In Mn doped ZnO, we find that PAL sensitive zinc vacancies/vacancy clusters at the grain surfaces contribute to enhance FM interaction. An effective way of controlling the related disorder is the suitable choice of mechanical milling and subsequent annealing.

A different interpretation of our PAL data is also possible. In nanomaterials, there exist abundant open volumes such as, interface junctions, triple junctions of the grains or several missing crystallites [37]. The third type of open volumes is larger in size and can be a source of positronium formation [25] and thereby giving rise to longer lifetime component, $\tau_3$. $\tau_2$ may originate from the open volumes associated with intergranullar spaces (second type). $\tau_1$ is associated with small size vacancy defects at the interface of grains (first type). As the defect density is very high in these nano-scale materials, one can expect saturation trapping [37] of positrons and so sensitivity of defect concentration by PAL is lost. Only the relative abundance of defect types (associated with $\tau_1$ or $\tau_2$) can be estimated from the parameter $I_1/I_2$, assuming that specific trapping rates of positrons at different defects remain unchanged. An increase of $I_1/I_2$ reflects the increase of weightage of the first type defects at the cost of intergranullar spaces in the sample. The variation of this parameter is shown in table 1. Increase of interface defects at the cost of intergranullar spaces may be due to the lowering of grain size distribution which is in qualitative agreement with the scanning electron micrographs of the samples (not shown). The change of $I_1/I_2$ with milling time is more convincing than that of $\tau_{av}$. An overall enhancement of $M_S$ with $I_1/I_2$ can be noted except for the 24 hour milled sample. Also the value of $\tau_1$ approaches towards 230 ps which is the experimentally measured lifetime of positrons at $V_{Zn}$. So improvement of ferromagnetic properties is connected with the abundance of $V_{Zn}$ in Mn doped ZnO.

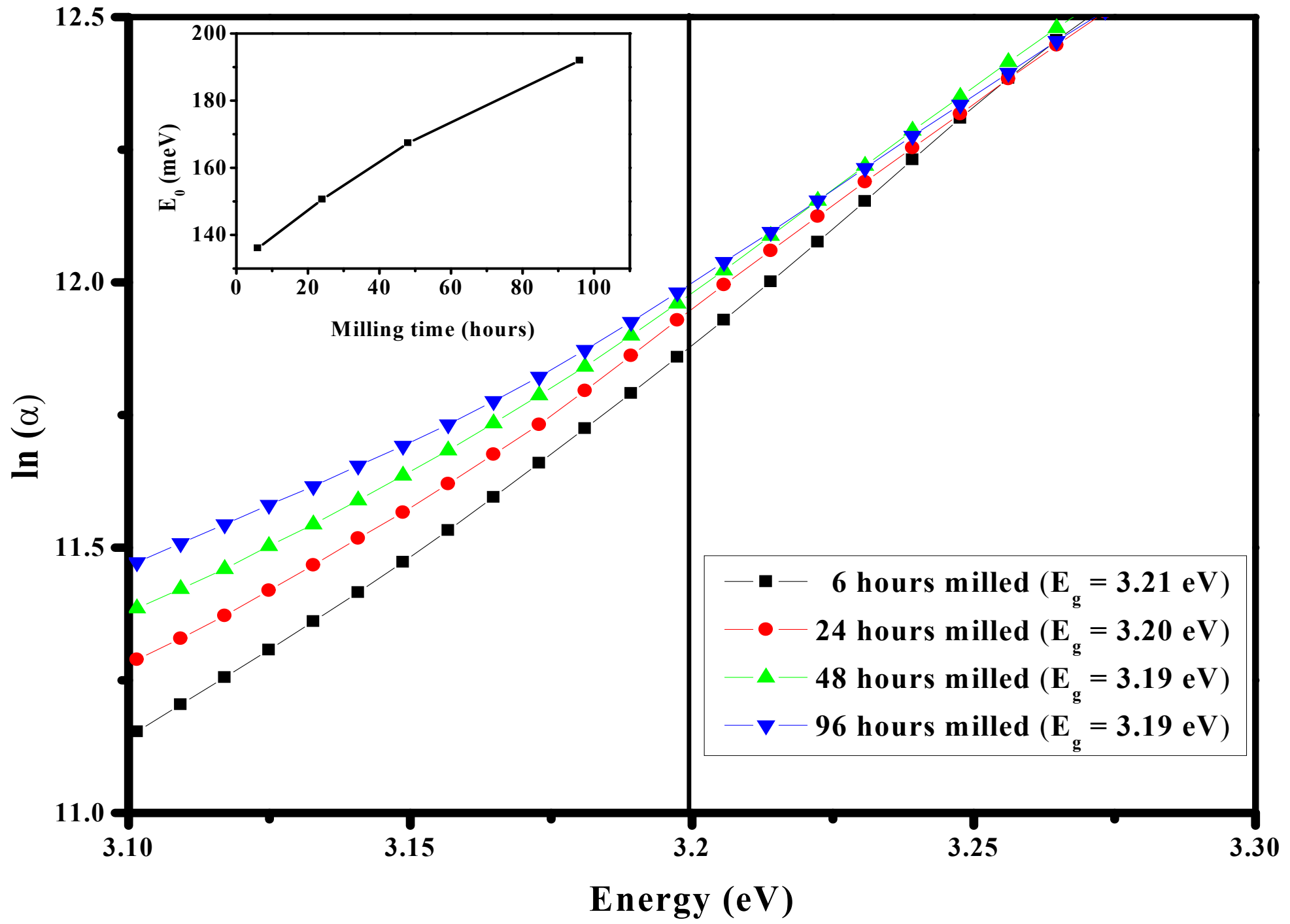

Figure 4: Plots of ln(α) with photon energy, where α is the absorption coefficient. The vertical line represents the mean band gap value. Inset: Variation of band tail parameter ($E_0$) with milling time.

Table 1: Physical parameters stemmed out of room temperature electrical resistivity and PAL measurements for doped ZnO samples milled for different hours.

| Resistivity (Ω-cm) | $\tau_1$ (ps) | $I_1$ (%) | $\tau_2$ (ps) | $I_1/I_2$ | $\tau_3$ (ps) |
|---|---|---|---|---|---|
| $4.64 \times 10^9$ | 171 ± 1 | 30.6 ± 0.1 | 329 ± 2 | 0.47 | 1358 ± 31 |
| $219.44 \times 10^9$ | 163 ± 1 | 22.3 ± 0.1 | 330 ± 2 | 0.30 | 1467 ± 32 |
| $250.58 \times 10^9$ | 190 ± 1 | 40.2 ± 0.1 | 350 ± 2 | 0.72 | 1427 ± 31 |
| $239.19 \times 10^9$ | 222 ± 1 | 59.3 ± 0.1 | 401 ± 2 | 1.58 | 1406 ± 48 |

We get further insight of the dominant presence of $V_{Zn}$ in the samples from analyzing the optical absorption data. The details of the measurement technique, data analysis and results have been published earlier [26,38]. Here, in Fig. 4 only the band tail region (lower energy part just below the band edge, $E_g$) have been shown. The band tail parameter ($E_0$), which reflects the defect concentration in the sample, is inverse the slope of the curves shown in Fig. 4. An increasing value of $E_0$ with higher milling time is clearly visible. The value of $E_0$ found here is also representative of acceptor defects [37] (mostly $V_{Zn}$ or $Mn_{Zn}$) present in the samples (effect of valance band tailing). Sufficient number of $V_O$ related defects in the system cause significant lowering of the band edge [38,39] (formation of deep centers). So the identification of zinc vacancies as key agent in inducing ferromagnetism in Mn doped ZnO is confirmed.

## 4. Conclusions

In summary, 2 at% Mn doped ZnO powder samples with different milling time have been synthesized by solid-state reaction technique. All the samples exhibit single-phase behavior and ferromagnetic at room temperature. PAL spectroscopy confirms the defect related origin of FM in the samples and the dominant defect is of zinc vacancy type. The value of the band tail parameter (extracted from UV-Visible absorption spectra) supports this contention. Modification of the defective state of the grain boundaries due to mechanical milling and subsequent annealing actually controls the observed ferromagnetic behavior of the samples. To achieve suitable defect configuration in this type of samples correct choice of milling time and annealing conditions is necessary.

## Acknowledgments

The authors thankfully acknowledge D. Jana, Department of Physics, University of Calcutta, for fruitful discussion. Financial assistance from DST-FIST, Government of India is also gratefully acknowledged. Author SKN is thankful to University Grants Commission (UGC) for providing him Junior Research Fellowship and author SC is grateful to Government of West Bengal for providing financial assistance in form of University Research Fellowship. The author SB is also thankful to Department of Science and Technology (DST), Govt. of India and IUAC, New Delhi for providing financial support in the form of research project vide project nos.: SF/FTP/PS-31/2006 and UFUP-43308 respectively.